\documentclass[]{aastex61}

\newcommand\aastex{AAS\TeX}

\received{July 1, 2016}
\revised{September 27, 2016}
\accepted{\today}
\submitjournal{ApJ}

\shorttitle{\aastex\ sample article}
\shortauthors{Centeno et al.}

\begin{document}

\title{Effects of spectral resolution on simple magnetic field diagnostics
  of the Mg {\sc II} h \& k lines }

\correspondingauthor{Rebecca Centeno}
\email{rce@ucar.edu}

\author[0000-0002-1327-1278]{Rebecca Centeno}
\affil{High Altitude Observatory (NCAR), 3080 Center Green Dr., Boulder, CO, USA}
\author[0000-0001-5850-3119]{Matthias Rempel}
\affil{High Altitude Observatory (NCAR), 3080 Center Green Dr., Boulder, CO, USA}
\author{Roberto Casini}
\affil{High Altitude Observatory (NCAR), 3080 Center Green Dr., Boulder, CO, USA}
\author{Tanaus\'u del Pino Alem\'an}
\affil{Instituto de Astrof\'\i sica de Canarias, C/ V\'\i a L\'actea
  S/N, 38204 La Laguna, Spain}

\begin{abstract}
We study the effects of finite spectral resolution on the magnetic
field values retrieved through the weak field approximation (WFA) from
the cores of the Mg {\sc ii} h\&k lines. The retrieval of the
line-of-sight (LOS) component of the magnetic field, $B_{\rm LOS}$,
from synthetic spectra generated in a uniformly magnetized FAL-C
atmosphere are accurate when restricted to the inner lobes of Stokes
V. As we degrade the spectral resolution, partial redistribution (PRD)
effects that more prominently affect the outer lobes of Stokes V, are
brought into the line core through spectral smearing, degrading the
accuracy of the WFA and resulting in an inference bias, which is more
pronounced the poorer the resolution. When applied to a diverse set of
spectra emerging from a sunspot simulation, we find a good accuracy in
the retrieved $B_{\rm LOS}$ when comparing it to the model value at
the height where the optical depth in the line core is unity. The
accuracy is preserved up to field strengths of B$\sim$1500 G. Limited
spectral resolution results in a small bias toward weaker retrieved fields. The WFA for the transverse component of the magnetic field is also evaluated. Reduced spectral resolution degrades the accuracy of the inferences because spectral mixing results in the line effectively probing deeper layers of the atmosphere.

\end{abstract}

\section{Introduction} \label{sec:intro}

Inference of the photospheric magnetic field vector has been
routinely carried out from ground-based observatories for decades.
With the advent of spectropolarimeters in space, these observations
(and their interpretation in terms of the vector magnetic field) have
reached the consistent (and continuous) high data quality required for
precise analysis of the magnetic field evolution. 
While the HMI instrument \citep[][]{HMI} on board the Solar Dynamics Observatory \citep[][]{SDO} provides full disk
measurements at good cadence and moderate
spatial resolution, the Hinode spectro-polarimeter \citep[][]{hinode, sp} offers a higher-resolution counterpart
within a much smaller FOV and limited temporal cadence.
The Solar Orbiter PHI instrument \citep[][]{PHI} will soon bring the first ever magnetic
field measurements at the solar poles with significant less
foreshortening than observations made from the ecliptic plane.
While each one of these observational benchmarks has furthered our
knowledge of the Sun's magnetism in a step-wise manner, the inferred
magnetic field is still confined to a thin atmospheric layer next to the Sun's
surface.

Between the dynamically driven photosphere and the
magnetically dominated corona, lies the ever-elusive chromosphere,
where the transition from the high- to low-$\beta$ regime takes
place. This interface region is also the connective tissue between the
Sun's surface and its outer envelope. The coronal heating conundrum
manifests indeed in the chromosphere, where the radiative losses due to
its proximity to outer space need to be compensated by efficient
heating mechanisms; and magnetic reconnection is a promising
candidate to explain said heating \citep[see, for instance, the discussion in][]{2017SSRv..210...37L}. 

Chromospheric magnetometry is challenging at many levels,
from the limits imposed by instrument sensitivity to the complexity of
the interpretation of the observations. Ground-based observations of magnetically-sensitive chromospheric
spectral lines (most notably He {\sc i} 10830 \AA, Ca {\sc ii} 8542 \AA\
and He {\sc i} D3 at 5876 \AA) have started to bridge the knowledge gap of what
happens in this interface region \citep[e.g.][among many
others!]{morosin2020, libbrecht2019, kleint2017, mjmg2015, spicules2010, merenda2006}.
\cite{jtb} make a case for spectropolarimetric observations in the
ultraviolet part of the spectrum for probing the magnetic field in the
upper chromosphere and the transition region.
Due to their high line core opacity as well as the relative brightness
of the UV spectral range where they sit, the Mg {\sc ii} resonance lines at 280 nm
are considered to be excellent probes, not only of the thermodynamics
\cite[][]{leenaarts1,leenaarts2}, but also the magnetic conditions in
the high chromosphere \cite[][]{hanlert, alsina_ballester, rafaSMM, judge2021}.

Leading up to the launch of the Interface Region Imaging Spectrograph
\cite[IRIS][]{IRIS}, a series of papers analyzed the properties and
modeling requirements of 
Mg {\sc ii} h\&k. \cite{leenaarts1} determined the minimum atomic model
necessary to reproduce the intensity spectrum of these lines, and studied the effects
of 1D vs 3D radiative transfer modeling, and including vs
ignoring the effects of partial frequency redistribution (PRD).
In a followup paper, \citet{leenaarts2} carried out a detailed study of
the morphological properties of Mg {\sc ii} h\&k intensity profiles
and their correlation with the thermodynamical properties in the
numerical simulation used to generate them. They found that the
positions and relative amplitudes of the spectral features serve as
proxies of the velocity and the temperature in the higher layers of
the chromosphere. Even after applying spectral and spatial smearing to
the synthetic data, some of these inferences remain robust \cite[][]{pereira2013}.
\cite{jaimeMg} show that non-LTE inversions of the spectral line intensity
allow for the retrieval of the full temperature stratification from
the mid-photosphere all the way
up to the transition region, whilst the velocity gradients can be
estimated throughout the chromosphere. This is a rather computationally
demanding approach, however, and for this reason,
\cite{IRIS2} developed a machine-learning approach that speeds up
traditional inversion methods by a factor of $10^5 - 10^6$, while
still reproducing the thermodynamical state of the
atmosphere with comparable quality to traditional minimization-based
techniques.

The IRIS observations, however, only measure the intensity of the Mg
{\sc ii} lines. Measuring the
spectral line polarization would open up the possibility of magnetic field
diagnostics in this crucial interface region at the base of the corona. In a renewed attempt to
analyze spectropolarimetric observations of Mg {\sc ii} resonance lines from the
Ultraviolet Spectrometer and Polarimeter on board the Solar
Maximum Mission \citep[SMM,][]{SMM}, \cite{rafaSMM} confirmed that the magnetograph formula
can be applied to observations of the circular polarization in the line
core. They obtained line-of-sight magnetic fields of the order of
500~G from these observations. The authors also report on observational evidence
of reduction of the broad-band linear polarization with respect to the
field-free case due to weak magnetic
fields outside of active regions. This had been theoretically predicted
by \cite{hanlert} and \cite{alsina_ballester} and presents a promising
diagnostic for the weaker chromospheric fields of the quiet Sun.

Decades after the SMM first measured the polarization of the Mg {\sc
  ii} UV doublet, the 2019 launch of the Chromospheric LAyer Spectropolarimeter
\cite[CLASP2,][]{CLASP2_WFA} revisited this part of the spectrum
with very promising findings. The circular polarization profiles of
h\&k in a plage region away from disk center were interpreted using the weak
field approximation (WFA) and revealed longitudinal magnetic
fields of up to $\sim300$~G at the top of the chromosphere, while a
nearby Mn I multiplet provided simultaneous information for lower layers.

Outside of active regions, the linear polarization in the core of Mg
{\sc ii} k (as well as the surrounding broad-band polarization)
is expected to be dominated by scattering phenomena. Trying to interpret these non-Zeeman signatures via the WFA
relations would lead to erroneous inferences.
Furthermore, most currently available non-LTE spectral line
inversion codes do not model the quantum-mechanical processes
responsible for scattering polarization and the Hanle effect,
also rendering them inadequate for the interpretation of these signals \citep[see][for an
analysis of this effect in Ca {\sc ii} 8542 \AA]{centeno2021}).
The newly published Tenerife Inversion Code
\citep[TIC,][]{TIC} is the first of its kind able to tackle
scattering polarization and Hanle-effect signatures in the Mg {\sc ii}
resonance lines. This is advancement will enable
detailed interpretation of the spectra, but comes at large
computational cost and can only be used sparingly\footnote{The computational expense of inverting a
  single spectrum is of the order of ~1000 core-hours.}.

While scattering polarization signals dominate the linear polarization
in and around the Mg II doublet \citep[except for the core of the h
line, see][]{hanlert, casini2002},
the core of Stokes V remains agnostic to these effects (del Pino
Alem\'an, private communication), allowing for a Zeeman
interpretation of the LOS component of the magnetic field. In the
presence of stronger fields, even the linear polarization becomes
dominated by the Zeeman effect, and retrieval methods based on it
should in principle still work.
In this paper, we study the error incurred by the WFA when applied to
the Mg {\sc ii} h \& k lines to obtain magnetic field inferences from
synthetic spectra emerging from an active region simulation. In particular, we look at how spectral
resolution degrades the accuracy of the results.
Section \ref{sec:tools} presents a brief overview of the
WFA as well as the radiative transfer code (Hanle-RT) that enables the spectral
synthesis. We first explore the effects of spectral resolution on the
WFA applied to
spectra emerging from a standard semi-empirical model atmosphere
(Section \ref{sec:FALC}), and then in Section \ref{sec:muram} we take the analysis to a more
diverse set of spectra emerging from a radiative magneto-hydrodynamic (rMHD)
simulation of an active region. We discuss the feasibility of the method and draw some concluding remarks in Section \ref{sec:conclusions}.

\section{Methodology}\label{sec:tools}
\subsection{The Weak Field Approximation}\label{sec:WFA}
The weak field approximation \citep[WFA; e.g.][]{landi_book} is a
simple, yet relatively robust method, to extract magnetic field
information from the intensity and polarization of spectral lines. As
its name suggests, its validity is limited to the ``weak'' field
regime, or more precisely, to the range of field strengths ($B$) in which the Zeeman-splitting
($\Delta\lambda_{\rm B}$) of the spectral
line is much smaller than its Doppler width, $\Delta\lambda_{\rm D}$:
\begin{equation}
\bar g \frac{\Delta\lambda_{\rm B}}{\Delta \lambda_{\rm D}} << 1 \label{eq:wfacondition}
\end{equation}

where $\bar g$ is the effective Land\'e factor of the spectral line and the
Zeeman splitting is proportional to the magnetic field strength ($B$,
in gauss) and the square of the central wavelength of the spectral
line under consideration ($\lambda_0$, in \AA):

\begin{equation}
\bar g \Delta\lambda_{\rm B} = 4.67 \times 10^{-13} \bar g \lambda_0^2 B 
\end{equation}

When applying a perturbative scheme to the radiative transfer
equation in this regime, one arrives at a series of relations that
connect the circular and the linear polarization spectra to
derivatives of the intensity with respect to wavelength.
Eq. \ref{eq:blos}  relates Stokes $V$ to the first derivative of
Stokes $I$ with respect to wavelength, through the line-of-sight (LOS)
component of the magnetic field vector, $B_{\rm LOS}$, and a constant $C_{\parallel}$
that depends on atomic properties of the spectral line.

\begin{equation}
  V(\lambda) = -C_{\parallel} B_{\rm LOS} \frac{\partial I}{\partial
    \lambda}
\label{eq:blos}
\end{equation}

\noindent Eqs. \ref{eq:bt1} and \ref{eq:bt2}, on the other hand,
relate the total
linear polarization ($L = \sqrt{Q^2+U^2}$) to the derivative of Stokes I through the square
of the
transverse component of the magnetic field vector, $B_{\rm T}^2$, and a
different constant, $C_{T}$, that also depends on properties intrinsic
to the spectral line:

\begin{equation}
L(\lambda_w) = \frac{3}{4}B_{T}^2
\left|C_{T}\frac{1}{\lambda_w-\lambda_0}\right|\left|\frac{\partial
    I}{\partial \lambda}\right|_{\lambda_w}\label{eq:bt1}
\end{equation}

\begin{equation}
  L(\lambda_0) = \frac{1}{4}B_{T}^2 \left|C_{T}\frac{\partial^2 I}{\partial \lambda^2}\right|_{\lambda_0}
\label{eq:bt2}
\end{equation}

where the first of these two equations is applicable only in the line
wing ($\lambda_w$) and the latter in the line core (close to the
central wavelength, $\lambda_0$).
The derivation of these relations as well as the conditions for their
applicability can be found in \citet{landi_book}.

The WFA has been
used in many works to extract quantitative information of the magnetic
field vector from spectropolarimetric observations \citep[see, for instance,][for
some recent applications of the WFA]{kriginski2021, vissers2021, siutapia,schukina}.
Chromospheric spectral lines tend to satisfy the WFA requirement of
Eq. \ref{eq:wfacondition} for
larger field strengths than their photospheric counterparts owing to
the fact that they are typically much wider than the latter. Additionally, lines that sit in the ultraviolet part of the
spectrum enjoy smaller Zeeman splittings due to their shorter
wavelengths, and lighter atomic species also meet the weak field
requirement better, since they typically experience larger Doppler broadenings
than their heavier counterparts.

The circular polarization of the Mg {\sc ii} resonance lines is known
to be amenable to the WFA \citep{hanlert}, and given strong enough fields, their
linear polarization signals may be interpreted with it as well.

\subsection{Spectral synthesis}\label{sec:hanlert}

Hanle-RT \citep[see][for a more comprehensive description]{hanlert} is a non-LTE polarized radiative transfer code for 1-D
(plane-parallel) stratified atmospheres. It accepts any multi-level or multi-term
atom (LS-coupling scheme) without hyperfine structure and in the
presence of magnetic fields of arbitrary strength. The code takes
into account the effects of non-isotropic radiative excitation of the
atomic system, which are responsible for the manifestation of
scattering polarization. Therefore, the code can model seamlessly all
possible regimes of magnetic-induced effects on the polarization of
spectral lines: the Hanle, Zeeman, and (incomplete) Paschen-Back
effects, as well as the polarization effects associated with level
crossing and anti-crossing interference. The code takes into account
the effects of inelastic and elastic collisions, and
also includes a comprehensive description of partially coherent
scattering (partial redistribution, or PRD) from polarized atoms
\citep[limited to 3-term atoms of the $\Lambda$-type;
see][]{2016ApJ...824..135C},
which is important for modeling deep chromospheric lines.

The atomic model of \ion{Mg}{2} used to solve the statistical
equilibrium (SE) includes 3 terms of the singly-ionized species as
well as the ground term of \ion{Mg}{3}. This is the smallest atomic
model necessary to model the atomic level polarization of the Mg {\sc
  ii} UV doublet around 280 nm as well as its subordinate triplet overlapping in the same wavelength
range.
The UV doublet is treated in PRD whilst the subordinate triplet is
computed assuming complete frequency redistribution.

\begin{figure}[!t]
  \includegraphics[angle=0,scale=.42]{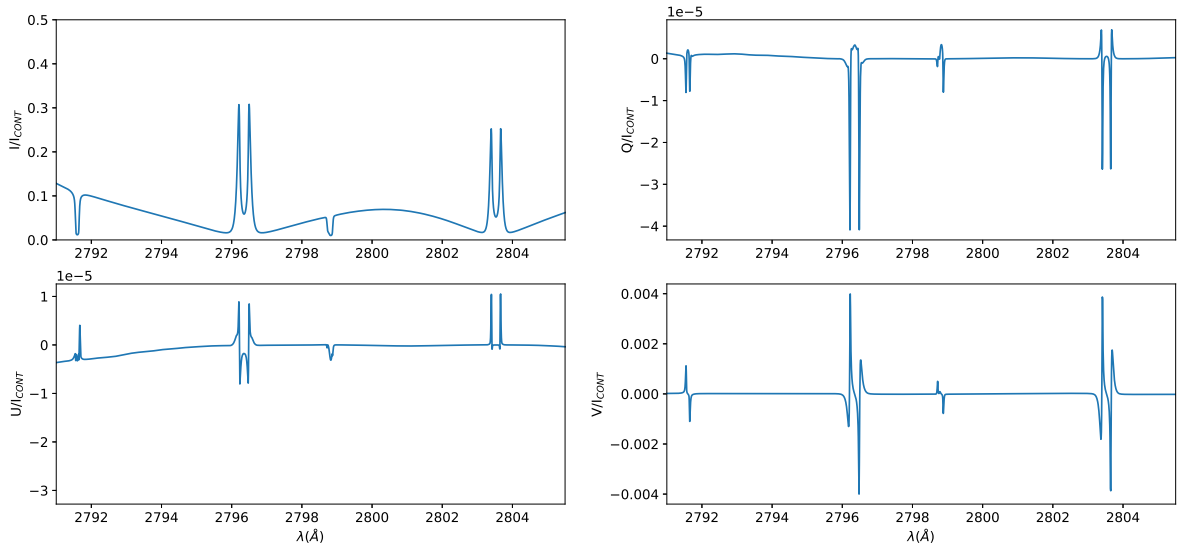}
\caption{Mg {\sc ii} h \& k as well as the subordinate UV triplet
  lines synthesized in a FAL-C model atmosphere with a magnetic field
  strength of $B=200$~G inclined $45^{\circ}$ from the local vertical,
  as if observed at disk center. Note the double lobes of the Stokes V
  profiles for the h \& k lines. The x-axis represents vacuum
  wavelengths, and Mg {\sc ii} k sits around 2796.3 \AA.
\label{fig:full_spectrum}}
\end{figure}

\section{Modeling in a FAL-C atmosphere}\label{sec:FALC}

We first synthesize the Mg {\sc ii} h \& k lines in the standard 1-D
FAL-C \citep[][]{FALC} semi-empirical model of the Sun's atmosphere
with an ad-hoc uniform magnetic field of $200$~G inclined $45^{\circ}$
with respect to the local solar vertical. For a disk center observing geometry,
this configuration results in a LOS component of the magnetic field of
$B_{\rm LOS} \approx 141$~G. A uniform magnetic field is chosen in order to
isolate the WFA inference error from other radiative transfer effects
resulting from LOS
integration. Fig. \ref{fig:full_spectrum} shows the synthetic spectrum
including the Mg {\sc ii} resonance lines as well as the subordinate triplet.
Three different syntheses are computed in this model atmosphere: the
first one is a full Hanle-RT calculation; in the second one we turn off magneto-optical effects; and in the
third one we turn off PRD effects instead.

We then apply Eq. \ref{eq:blos} by carrying out a linear regression ($y_i = a \cdot x_i + b$) of
the data to determine the slope $a$ and the intercept $b$. Here, $y_i$
correspond to the values of $V_{\lambda}$, while $x_i$ are taken to be
the values of
$-C_{\parallel} dI_{\lambda}/d \lambda$, with the proportionality
between the two given by the fitted slope ($a = B_{\rm{LOS}}$), and a fitted
intercept $b$ that should be compatible with zero. Because the WFA is
just an approximation, the linear relationship between $V_\lambda$ and
$dI_\lambda / d \lambda$ does not hold outside of the Zeeman weak field
assumption. When the data don’t meet the criteria for the WFA (and the
relationship is no longer linear), the linear fit might yield an
intercept that is not compatible with zero, and a slope that is
wrongly interpreted as $B_{\rm LOS}$.

\begin{figure}[!t]
  \includegraphics[angle=0,scale=.42]{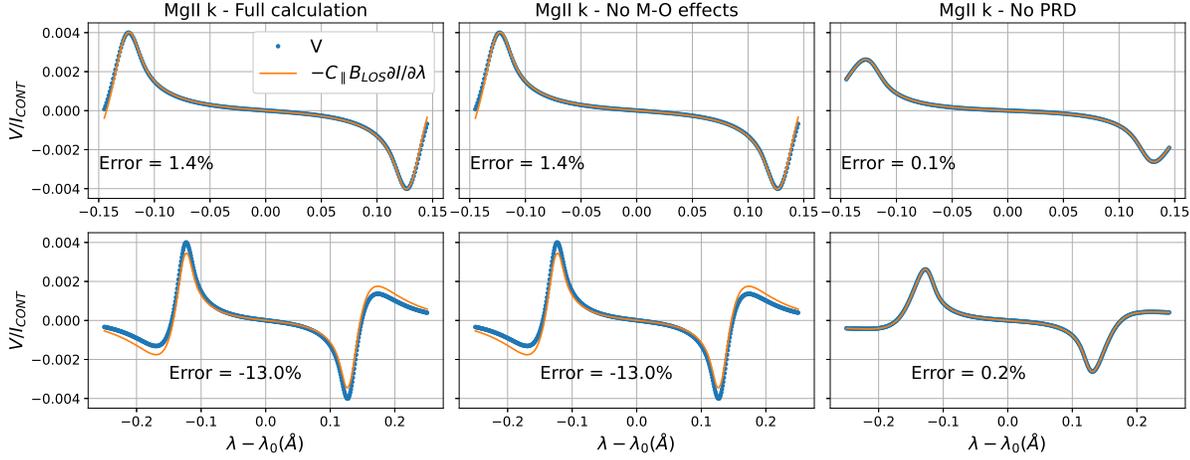}
\caption{WFA fit of Stokes V for Mg {\sc ii} k for the cases where the
  spectra were synthesized with no approximations (left), when
  magneto-optical effects were turned off (middle) and when no PRD
  effects were considered (right). The top row corresponds to the WFA
  applied on the inner lobes of Stokes V only, whilst the bottom row
  applies the equation to both sets of lobes. The error incurred by
  the WFA is reported inside each panel.
\label{fig:inner_outer_lobes}}
\end{figure}

Fig. \ref{fig:inner_outer_lobes} shows the WFA fit for the LOS component
of the magnetic field applied to the three synthetic profiles of Mg
{\sc ii} k. The blue dots represent the values of Stokes V (rearranged as a function
of increasing wavelength and normalized by the
local continuum intensity), while the orange line shows the derivative of
Stokes I with respect to wavelength, scaled by $C_{\parallel} 
B_{\rm LOS}$, where $B_{\rm LOS}$ was obtained from a linear regression
applied to Eq. \ref{eq:blos}. The relative error incurred by the WFA is
reported for each case inside the corresponding panel. In
the case where the WFA is applied to the inner lobes of Stokes V only
(top row) the inference error is $1.4\%$ when the line is treated in
PRD (left and middle panels), compared to 0.1\% in the scenario where
PRD is turned off (right panel). However, when the outer lobes are
taken into account in the linear regression (bottom row), the inference errors jump to
$-13\%$ in the two PRD cases (bottom left and middle panels), which is
also evidenced by the poor quality of the fit. When PRD is turned off, the error is
still very small (bottom right) and the retrieved $B_{\rm LOS}$
produces a good fit. This shows that the inaccuracy incurred by the WFA is
mostly due to PRD effects and can be avoided by excluding the outer
Stokes V lobes from the calculation. On the other hand,
magneto-optical effects alone (for this regime of field strengths) do not have negative
impacts on the WFA retrieval of $B_{\rm LOS}$\footnote{according to \cite{landi_book}, M-O
effects yield corrections of the order of
$(\Delta\lambda_B/\Delta\lambda_D)^4$ for the derivation of the LOS
magnetic field from the WFA (pp. 488-489).}.

In CRD, the WFA corresponds to the lowest-order
result of a perturbative solution of the polarized radiative transfer
equation in the presence of a magnetic field \citep[][]{Landi1973}. In
PRD, however, the emissivity vector $\epsilon$ gets modified by a
redistribution correction $\delta\epsilon$ \citep[see, e.g.][]{casini2017},
which breaks the symmetry of such a perturbative solution, and is
ultimately responsible for the observed departure of the validity of
the WFA in PRD away from the line core. We believe this is the only
plausible physical mechanism that can be responsible for the WFA
breaking down away from the line core.

\subsection{Effects of spectral resolution}

Most instruments that observe solar spectral lines have limited
spectral resolution. Spectral smearing mixes information across the
wavelength domain, possibly bringing information
from the line wings into the line core, and vice-versa. For the WFA, this means that
information captured by the line wings (which ``sense'' magnetic fields at lower
atmospheric heights) may be dragged into the core of the line, resulting
in line-of-sight smearing of the retrieved magnetic quantities.
But even when the magnetic field is constant with height, the PRD
effects that are prominent in the outer lobes of Stokes V can be
partially transferred to the inner lobes, and ``contaminate'' the WFA
inference applied to the narrower spectral range.

In order to quantify the latter effect, we apply gaussian point spread functions (PSF) of
different widths to the synthetic spectra described above, and apply
the WFA to the inner lobes of Stokes V  in the degraded
spectra. Because the spectral syntheses were carried out in an
atmosphere with a uniform magnetic field, the line-of-sight smearing
of the retrieved magnetic properties is inconsequential.

\begin{figure}[!t]
  \includegraphics[angle=0,scale=.4]{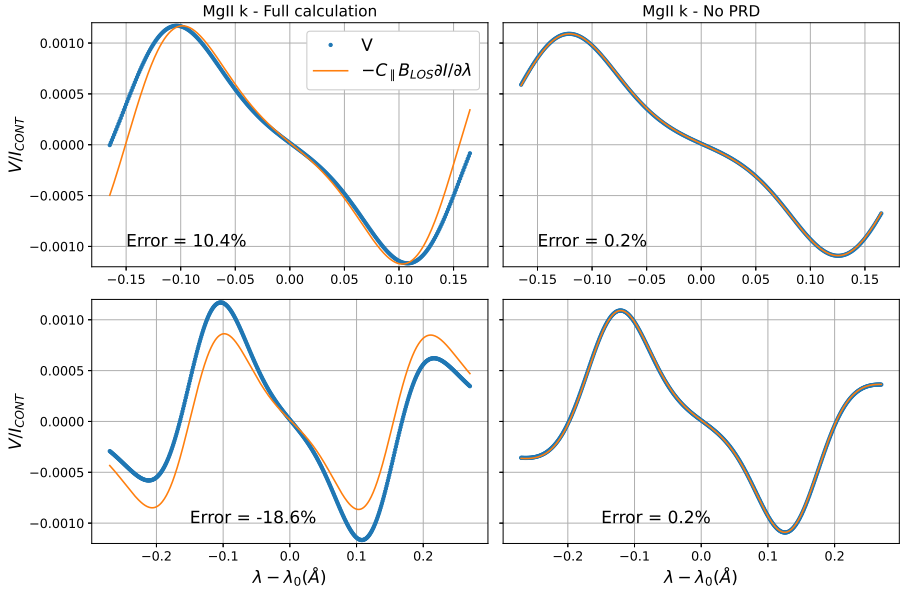}
  \caption{Application of the WFA fit (orange) to Mg {\sc ii} k Stokes V
    spectra (blue) that have been previously degraded with a gaussian point
    spread function corresponding to a spectral resolution of R =
    30,000. In the top row, the WFA has been applied to the inner
    lobes of Stokes V only, whilst the bottom panels show the results
    when applied to a wider wavelength range. PRD effects in the full
    calculation (left) result in large inference errors that are not
    present when PRD is turned off (right).
\label{fig:R30K}}
\end{figure}

The top left panel of Fig. \ref{fig:R30K} shows the WFA fit (orange lines) for Mg {\sc ii} k
applied to the inner lobes of Stokes V (blue dots) in the case of the
full calculation with PRD and a spectral
resolution of R = 30,000 (which is approximately the spectral resolution of the
CLASP II observations). The relative error in the retrieved $B_{\rm
  LOS}$ is $\sim 10.4\%$. Including the
outer Stokes V lobes in the WFA inference (bottom left) drives
the relative inference error to $-18\%$. On the other hand, when PRD is turned off (right panels), the inference error
remains very low ($0.2\%$) in both cases.
This example epitomizes
how the PRD effects that are prominent in the outer lobes, are brought into
the inner lobes due to spectral smearing, negatively impacting the WFA
inference.

\begin{figure}[!t]
  \includegraphics[angle=0,scale=.6]{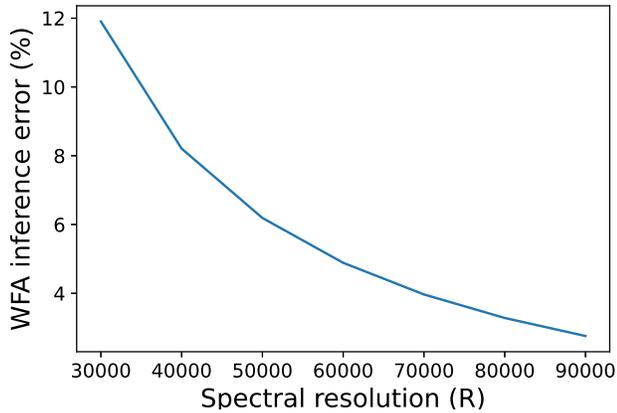}
  \caption{Error incurred by the WFA when inferring for $B_{\rm LOS}$ from the Mg
    {\sc ii} k line as a function of spectral resolution. The WFA
    was applied to the inner lobes of Stokes V only.
\label{fig:WFA_with_R}}
\end{figure}

Fig. \ref{fig:WFA_with_R} shows the relative error of the WFA results
for $B_{\rm LOS}$ as a function of spectral resolution of the
observations. As expected, smaller values of R result in greater
retrieval errors. It is important to note that, without the effects
of PRD, spectral convolution alone should (and does) not impact the
performance of the WFA in the case of a uniform magnetic field.

\section{The WFA applied to synthetic Mg {\sc ii} spectra
  emerging from an active region simulation}\label{sec:muram}
\begin{figure}[!t]
  \includegraphics[angle=0,width=0.98\textwidth]{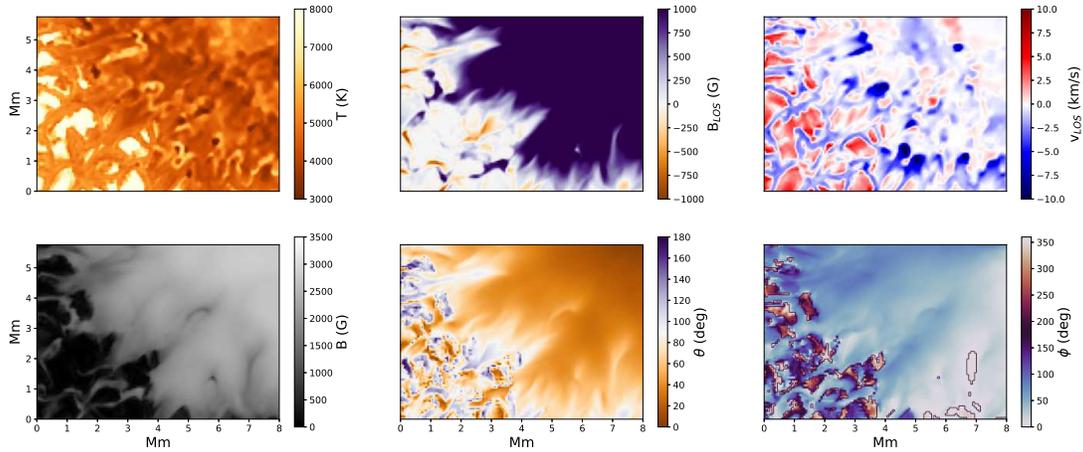}
  \caption{Physical parameters at h=0~km in the subset of the original
    MURaM simulation chosen for this study. From
left to right and top to bottom: temperature, $T$, LOS
component of the magnetic field, $B_{\rm LOS}$, LOS velocity, $v_{\rm
  LOS}$, magnetic field strength, $B$, its inclination with
respect to the local vertical, $\theta$, and its azimuth in the horizontal
plane, $\phi$).
\label{fig:simulation}}
\end{figure}

In this section we analyze the diagnostic capabilities of the WFA when
applied to Mg {\sc ii} lines emerging from an active region
simulation, and in particular, how the accuracy of the WFA is affected by finite
spectral resolution.

We start from a radiative magneto-hydrodynamic (rMHD) simulation of
an active region (AR) from the MURaM code with the same setup as the
one described in
\cite{rempel_MURaM}. 
The simulation presents a small active region with a corona that is
self-consistently heated by the Poynting flux resulting from
photospheric magneto-convection. While the treatment of the
chromosphere is simplified (grey RT and LTE ionization),
\cite{bjorgen} showed
that synthetic chromospheric emission captures many observationally
known features, at least on a qualitative level. The purpose of this
simulation is not to provide an exact description of an active region
chromosphere, but to provide a test case with known "ground truth"
that samples the typical parameter space of an active region.
The size of the horizontal domain is
98x98 Mm$^2$ with a spatial sampling of 64 km in each horizontal
direction. The vertical domain spans $\sim 49.1$~Mm (sampled every 32 km), of
which approximately $\sim 13.1$~Mm lie below the surface (defined as the height
where the average optical depth at 500~nm, $\tau_{500}$, is unity).
We select a small subset of the simulation box that encompasses a quarter of one of
the sunspots and the strongly magnetized granulation surrounding it to synthesize the Mg {\sc
  ii} lines in the UV. Fig. \ref{fig:simulation} presents the physical
variables of the simulation in the horizontal slice at h=0~km. The strong and almost vertical magnetic field at the
center of the umbra (top right corner of each panel) decreases rapidly in strength towards the outer edge of the
sunspot, where it also becomes almost horizontal.

\begin{figure}[!ht]
  \includegraphics[angle=0,width=0.5\textwidth]{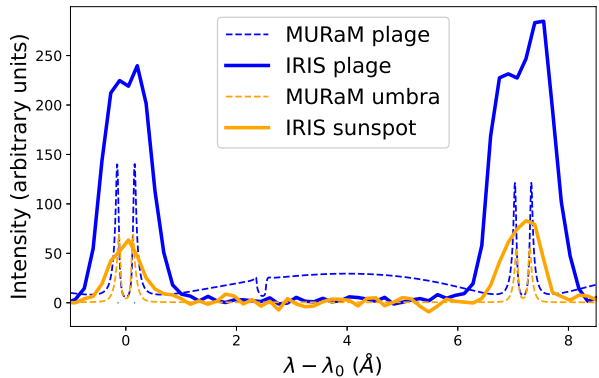}
  \caption{Dashed lines: averaged synthetic Mg {\sc ii} spectra over a 240x240 km$^2$
    area in the top right (umbra) and bottom left (``plage'') corners of the
    simulation box. For comparison, Level 2 IRIS spectra inside a
    sunspot and a plage area are shown with solid lines. Note that the
    individual synthetic spectra averaged for this figure were
    calculated at rest (no Doppler shift),
    and were not subjected to 
    spectral degradation. Both of these factors contribute to them
    being narrower than the observed spectra. In this figure, the
    MURaM and the IRIS specific intensities were
    scaled arbitrarily.
\label{fig:synthetic_vs_iris}}
\end{figure}

We extract the physical quantities for each column (1D atmosphere) in
the simulation and feed it to Hanle-RT to generate the Stokes
spectra. Before feeding the atmospheres to Hanle-RT some pre-processing is
necessary. Each column of the simulation is truncated 192~km
  below its $\tau_{500}=1$ level. 
The relative shift between this level across individual
atmospheres in the simulation domain can be as large as 800km due to
the surface being
very depressed at the center of the umbra compared to the surrounding
granulation.
The top of each atmosphere is also truncated at the location where the temperature
reaches $10^5$K, since the Mg {\sc ii} lines should form well below said
temperature \citep{heinzel2014}.
The atmospheres are further simplified by removing the velocity field.
Because there are no velocity gradients, the
emerging spectra will be close to symmetric. This simplification is
necessary for the purpose of isolating the effects of spectral resolution on the WFA
inferences from further vertical smearing that the velocity gradients
may introduce \cite[although][showed that moderate velocity gradients do not
negatively impact WFA inferences of $B_{\rm LOS}$ from Ca {\sc ii}
8542 \AA]{centenoWFA}.
Furthermore, an ad-hoc microturbulent velocity profile is added to
each column's atmosphere. The latter is modeled as a linear function of height, ranging from
0~kms$^{-1}$ at h=-192 km, to 7~kms$^{-1}$ at the highest possible
height (h=1856~km), so that it has qualitatively comparable values to
those in the semi-empirical model atmosphere FALC, as well as in the
inversion results of sunspot/plage IRIS spectra in
\cite{jaimeMg}. This high microturbulence value still produces spectra
that are much narrower than those in Level 2 IRIS data, as shown in Fig \ref{fig:synthetic_vs_iris}.

The spectra are synthesized in the 3-term Mg {\sc ii} atom described
above, for a disk center observing geometry (i.e. a heliocentric angle
of $0^{\circ}$).
Synthesizing the Mg {\sc ii} h\&k lines plus the UV triplet in the
3-term atomic model with Hanle-RT is a computationally demanding
exercise. The calculation of the atomic level polarization (population
imbalances and quantum coherences amongst magnetic sublevels) as well
as the treatment of the Mg {\sc ii} h\&k doublet in PRD dramatically
drive the increase in computing time compared to spectral line
modeling without these effects.
The median run time for generating a single spectrum is 10 core-hours (c-h), and synthesizing the
11,250 spectra in the FoV in Fig. \ref{fig:simulation} took a
total 145,000 c-h on NCAR's Cheyenne supercomputer.

In the case of sunspots,
magnetic fields are strong enough to significantly depolarize the atomic
levels, leading to polarization signals that exhibit, almost typical,
Zeeman patterns. In this regime, if the spectral lines are wide
enough, the WFA can be a suitable tool for quantifying the magnetic
field vector because the inference error induced by any residual
atomic polarization can typically be ignored for the expected field
strengths of active regions \citep{kuckein}.

\subsection{Retrieval of B$_{\rm LOS}$ from Mg {\sc ii} h \& k} 

In this section we apply Eq. \ref{eq:blos} to retrieve the LOS component
of the magnetic field from Stokes I and V of the Mg {\sc ii} h
and Mg {\sc ii} k lines separately, but first, we implement an automated algorithm that finds the wavelength range
corresponding to the inner lobes of Stokes V based on the position of
its peaks (which appear as zero-crossings in the first derivative of
Stokes V with respect to wavelength). Even in
the absence of velocities, the shape of the spectral line changes
significantly across the FoV, and a single pre-chosen wavelength
interval is not adequate to exclude the outer Stokes
V lobes in all the spectra. 

\begin{figure}[!t]
  \includegraphics[angle=0,width=0.98\textwidth]{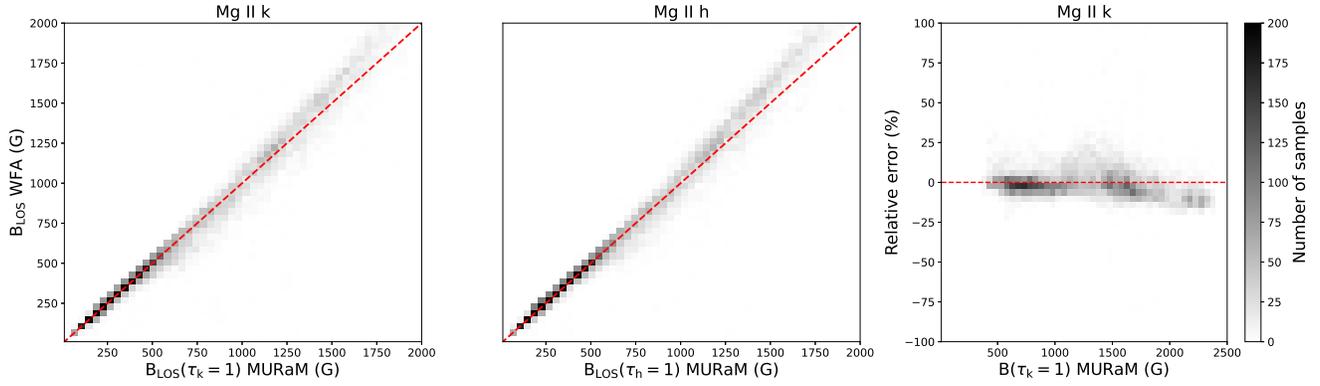}
  \caption{Scatter density plots of the retrieved value of B$_{\rm
      LOS}$ (y-axis) against their model counterparts where
    $\tau_{\rm LINE}=1$ (x-axis) for Mg {\sc ii} k (left) and h (middle). The darker the grey-level, the higher the number of
    samples in the bin. The bin size is 40 G. The red line represents the
    ideal solution, if the WFA retrieved exactly the model value. The
    panel on the right shows the relative error incurred by the WFA
    applied to Mg {\sc ii} k as a function of the magnetic field
    strength in the MURaM model.
\label{fig:WFA_Blos_hvsk}}
\end{figure}

\begin{figure}[!t]
  \includegraphics[angle=0,width=0.5\textwidth]{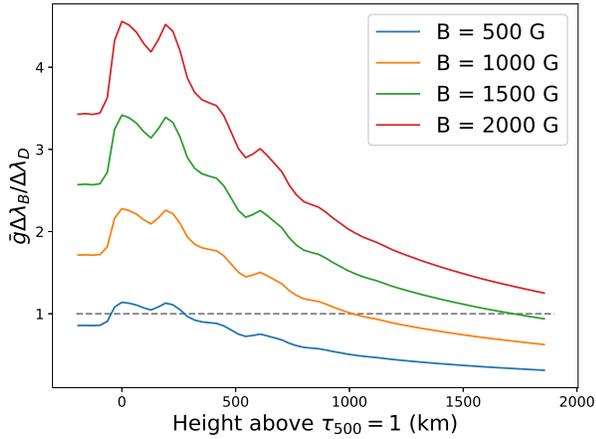}
  \caption{Meeting the WFA condition. The lines show the value of
    $\bar g \Delta\lambda_B/\Delta\lambda_D$ for the Mg {\sc ii} k line in one of the MURaM atmospheres,
    where the magnetic field has been replaced by a uniform one a
    function of height. Three different magnetic strength values are
    considered, showing that, at the typical height of formation of
    the Mg {\sc ii} lines, a field of 1500~G no longer meets the
    condition of Eq. \ref{eq:wfacondition}.
\label{fig:DeltaLambdaB}}
\end{figure}

After finding the desired wavelength interval, we carry out a linear fit between
the left- and the right-handsides of Eq. \ref{eq:blos}, in the same
way as in Section \ref{sec:FALC}, for each set of Stokes $I$ and $V$
emerging from the FoV of the simulation box.
Fig. \ref{fig:WFA_Blos_hvsk} shows density scatter plots of the retrieved
$B_{\rm LOS}$ from Mg {\sc ii} k (left) and Mg {\sc ii} h (middle)
against the MURaM model values. The latter are taken at the height where each line center reaches optical
depth unity ($\tau_{\rm h}$ or $\tau_{\rm k}$ = 1). This is taken
  as a representative height for comparison with the model values,
  however, the emergent spectra are sensitive to a range of heights
  over which the magnetic field is not constant. The gradient of the
  magnetic field around the region of formation of the lines is
  responsible for the scatter in the WFA inferences. 
The magnetic field bin size is 40~G, and the darkness
of the gray-scale is proportional to the number of samples in the
bin. The red line represents the ideal solution (if the WFA yielded
exactly the model value). From these figures it is clear that the WFA does a good
job at retrieving the LOS component of the magnetic field up to LOS field strengths of $B_{\rm LOS}\sim 1000$~G. Beyond this value, the WFA
starts to systematically overestimate $B_{\rm LOS}$. This is 
due the fact that we are entering the strong field regime in which the
WFA is no longer valid, according to the requirement of
Eq. \ref{eq:wfacondition}. 
Note that the validity of the weak field regime hinges on the value of
$B$, not of $B_{\rm LOS}$. A LOS component of 1000~G necessarily
entails a field strength larger or equal to that. The right
panel of Fig. \ref{fig:WFA_Blos_hvsk} presents the relative error incurred by
the WFA when applied to the Mg {\sc ii} k line as a function of the
total field strength, B, in the
model. This panel clearly shows that the systematic
deviation of the WFA results towards larger values starts around
$B\sim 1500$~G.
 Fig. \ref{fig:DeltaLambdaB} shows the values of
  $\bar g\Delta\lambda_B/\Delta\lambda_D$ for the Mg {\sc ii} k line as a
  function of height in one of the 1D MURaM
  atmospheres, for four different (uniform) magnetic field strengths.
  Around 1500~G, the Mg {\sc ii} k line no longer satisfies the WFA
  condition. This threshold varies from one atmosphere to another,
  since meeting or not the condition depends on the temperature
  and the microturbulent velocity as well as the magnetic field strength.

\subsection{Effects of spectral resolution on the retrieval of B$_{\rm
    LOS}$ }
\begin{figure}[!t]
  \includegraphics[angle=0,width=0.98\textwidth]{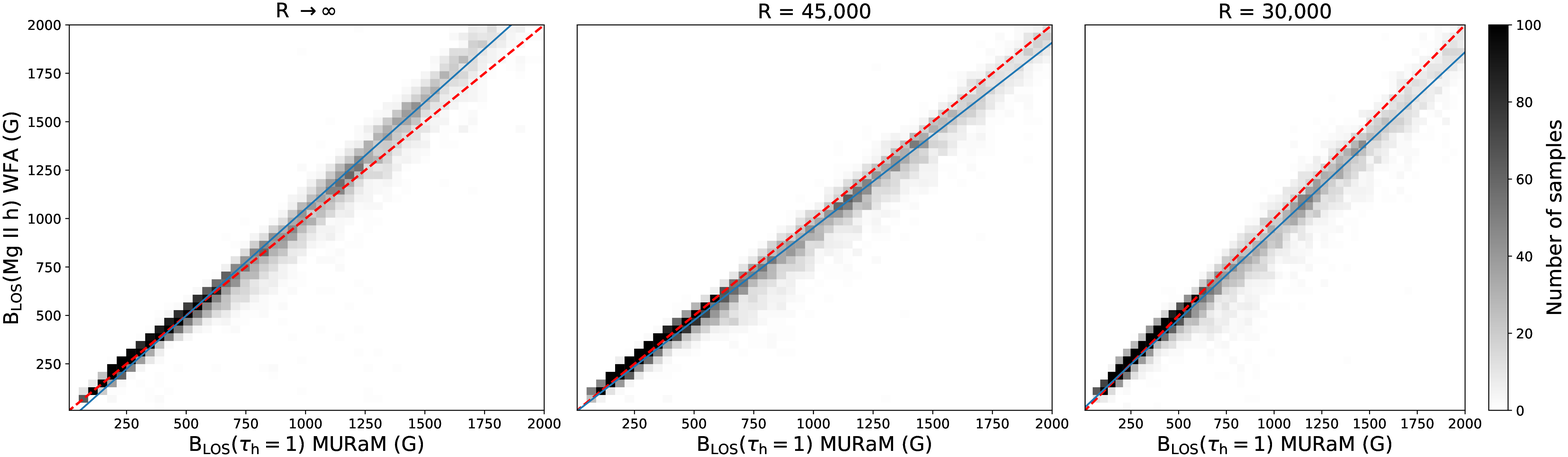}
  \caption{Scatter density plots of the retrieved value of B$_{\rm
      LOS}$ retrieved from Mg {\sc ii} h against their model
    counterparts for the case of infinite spectral
    resolution (left), R = 45,000h (middle) and R = 30,000 (right). The
    model values are taken at the height where the core of each line
    reaches optical depth unity. The darker the grey-level, the higher the number of
    samples in the bin. The bin size is 40 G. The red line represents the
    ideal solution and the blue line shows a linear fit through the
    data. The left panel in this figure shows the same data than the
    middle panel of Fig. \ref{fig:WFA_Blos_hvsk}
\label{fig:WFA_Blos_vs_R}}
\end{figure}
In order to understand the effects of a finite spectral resolution on
the WFA inferences, we convolve the spectra emerging from the simulation with gaussian
PSFs of different widths. 

Fig. \ref{fig:WFA_Blos_vs_R} shows scatter plots of the WFA
inference against the corresponding model value for the cases of
``infinite'' spectral resolution (left),  R = $\Delta\lambda/\lambda =$45,000 (middle) and  R =
30,000 (right). The red line represents the one-to-one solution while
the blue line shows a linear regression of the data points in each scatter plot. As discussed earlier,
the WFA tends to over-estimate the field above $B_{\rm LOS}=1000$~G in
the spectrally resolved case (left panel). As the value of $R$
decreases, the method starts to underestimate $B_{\rm LOS}$,
counteracting the effect of the WFA inaccuracy in the strong field regime.
This systematic bias towards lower fields (seen in the middle and right panels) cannot be
attributed to the core of the line sensing lower atmospheric
layers due to spectral smearing (i.e. along the LOS) mixing. Spectral information probing lower atmospheric layers would
typically lead to a bias towards larger inferred field strengths, not
weaker, because there is an almost monotonic decrease of the magnetic field strength as a function of
height across the entire FoV of Fig. \ref{fig:simulation}. This points
to PRD effects being brought into the core of the line due to spectral smearing
and thus impacting (decreasing) the WFA inferences, which should
otherwise increase with spectral mixing (i.e. mixing
along the line-of-sight). The larger the value of R, 
the greater the impact on the inferred $B_{\rm LOS}$.

\subsection{Retrieval of B$_{\rm T}$ through the WFA}\label{sec:bt}
The retrieval of the transverse magnetic field through the WFA relies
on Eqs. \ref{eq:bt1} and \ref{eq:bt2}.
While Eq. \ref{eq:bt1} is applicable far
away from the central wavelength of the spectral line, $\lambda_0$, Eq. \ref{eq:bt2} is
technically valid only in the core. For this reason alone, the former
can only be used to sample lower atmospheric layers from the wings of the
Mg {\sc ii} lines, whilst the latter should be able to probe the
transverse component of the magnetic field in the
higher layers of the atmosphere. A practical caveat to both relations is
that, when the magnetic field is very weak, the linear polarization
signals tend to be dominated by scattering polarization, the Hanle
effect, and magneto-optical effects, rendering the WFA a useless tool in these cases.
The magnetic field needs to reach a few hundred gauss in
strength in order to significantly depolarize the atomic energy levels of the
upper level of the Mg {\sc ii} UV doublet, and allow the linear
polarization signals to enter the Zeeman-dominated
  regime, where the WFA is valid.
  In the case at hand, the weakest magnetic field in the entire FoV is still
  $\approx 500$~G at the top of the atmosphere\footnote{The top of the
  atmosphere in this context is defined by the height at which the
  temperature reaches $10^5$~K}.

 Since the retrieval of the magnetic field at the top of the
  chromosphere is of most interest to this exercise, we will focus on
  testing the accuracy of the WFA through Eq. \ref{eq:bt2}.
Three different applications of this equation are tested within the core of the line, here defined as the
wavelength range occupied by the inner lobes of Stokes V. Method \#1 performs a standard
linear regression of the left and right handsides of Eq. \ref{eq:bt2}. In the
linear regression analogy described in Section \ref{sec:FALC}, $y_i$
now corresponds to $L_\lambda$, while
$x_i$ takes the values of $1/4 C_T |d^2I_\lambda / d \lambda^2|$. The slope of
the fit corresponds to $B_{\rm T}^2$, while the intercept should be
close to zero if the model represents the data adequately.
In method \#2, the intercept $b$ is forced to be zero and only the
slope is fitted, which results in the following expression for the linear regression:
\begin{equation}
B_{\rm T}^2 = \frac{4}{C_{\rm T}}\frac{\sum_\lambda(\big|\frac{\partial^2
    I_\lambda}{\partial\lambda^2}\big| L_\lambda)}{\sum_\lambda \big(
  \frac{\partial^2
    I_\lambda}{\partial\lambda^2}\big)^2}
\end{equation}

 In method \#3, we calculate the wavelength-average of the left- and
 the right-hand sides of Eq. \ref{eq:bt2} and obtain $B_{\rm T}^2$ as the
 ratio between the two:
 \begin{equation}
   B_{\rm T}^2 = \frac{4}{C_{\rm T}} \frac{\sum_\lambda L_\lambda}{ \sum_\lambda \big|\frac{d^2I_\lambda}{ d \lambda^2}\big|}
\end{equation}
 
 Additionally, we test the application of Eq. 13 in
 \cite{marian2012} (method \#4). This approach, estimates the magnetic field vector
 by minimizing a merit function that explicitly uses Stokes $Q$
 and $U$, instead of $L$. This results in the following equation to
 obtain $B_T$:

 \begin{equation}
B_T^2 =\frac{4}{C_{\rm T}}
\frac{\sqrt{\sum_{\lambda}(Q_{\lambda}\frac{\partial^2I_{\lambda}}{\partial\lambda^2})^2
  + \sum_{\lambda}(U_{\lambda}\frac{\partial^2I_{\lambda}}{\partial\lambda^2})^2}}{\sum_{\lambda}(\frac{\partial^2I_{\lambda}}{\partial\lambda^2})^2}\label{eq:marian}
   \end{equation}

\begin{figure}[!t]
  \includegraphics[angle=0,width=0.98\textwidth]{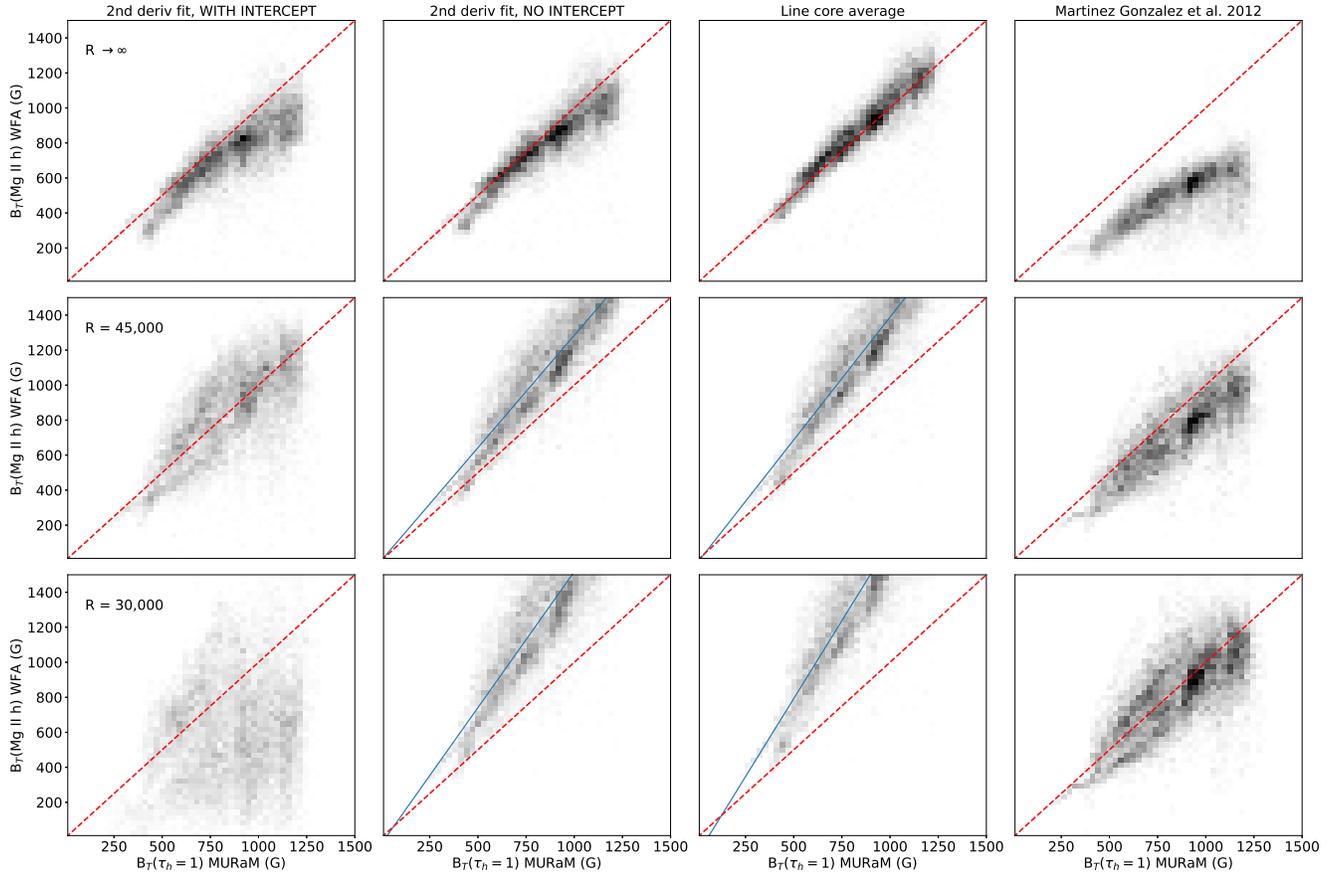}
  \caption{Scatter density plots of the values of B$_{\rm
      T}$ retrieved from Mg {\sc ii} h against their model
    counterparts for the case of infinite spectral
    resolution (top row), R = 45,000h (middle row) and R = 30,00
    (bottom row). The four columns correspond to retrieval methods \#1, \#2, \#3
    and \#4 respectively, described in the manuscript text. The darker the grey-level, the higher the number of
    samples in the bin. The bin size is 40 G and the color scale is
    the same for all panels. The red line represents the
    ideal solution if the WFA inference were perfect, while blue lines
    show linear fits through the scatter plot data. 
\label{fig:WFA_Btrans_vs_R}}
\end{figure}

Fig. \ref{fig:WFA_Btrans_vs_R} shows the results of applying
the WFA to Mg {\sc ii} h for inferring the transverse component of the field from the synthetic
spectra using these four different methods (from left to right, the
columns correspond to methods \#1, \#2, \#3 and \#4, respectively). 
Each panel presents a density scatter plot of the
retrieved transverse magnetic field against its model counterpart at
the height where $\tau_{\rm Mg II h }=1$. The
three rows show the results for different spectral
resolutions, R.

Let's take a look at the first row of results, for the spectrally resolved case. 
The first and second panels show the WFA results from the linear regressions of Eq. \ref{eq:bt2}
with and without intercept, respectively (methods \#1 and \#2). We find that, fixing the
intercept to zero yields slightly more accurate results than allowing it to be
a free parameter in the regression. This is a symptom of the WFA model
not representing the data accurately (i.e. the best linear fit does not naturally cross the origin).
The third panel shows the case where the linear fit is
forgone and only a ratio of the wavelength-averaged quantities is
used. The accuracy is comparable to that of the previous 
methods. The last panel shows the results from the application of
Eq. \ref{eq:marian}, which systematically underestimates the values of
$B_T$.

As we decrease the spectral resolution to 45,000 (middle row) and
30,000 (bottom row) we find that the accuracy of all methods degrades, with method \#1 losing all correlation and retaining no diagnostic value at
R = 30,000.
Methods \#2 and \#3 show very similar behaviors as the spectral resolution
worsens. They both result in an increased spread of the WFA solution
around the model value, and they over-estimate the transverse field by a larger amount the
poorer the resolution. The spectral smearing is
bringing information from the line wings into the core and
``contaminating'' the WFA inferences. This over-estimate is partly due
to the fact that the wings of the line sense lower layers (and therefore stronger
fields). The impacts of PRD effects and scattering polarization in the
line wings
could also be contributing to the error. From the setup of
this particular experiment it is impossible to quantify the relative
contributions of each one of these effects. In any case, this over-estimate is relatively well behaved (systematic as
a function of field strength) and could be
corrected with a proper calibration of the WFA inferences
\footnote{This may be MHD model-dependent, in which case may only be
  done when MHD models are able to reproduce the observed spectra.}. The blue
lines in these panels show a linear fit through the scatter
plot data, suggesting that a simple linear correction to the WFA
inference could bring it closer to the real $B_{\rm T}$ value.
Method \#4 also results in an increased spread with systematically larger inferred values as we degrade the
resolution.

It is worth noting that, while the amplitudes of the linear
polarization signals emerging from the AR simulation reach values of the order of 1\% of the
continuum intensity, we might expect them to be much smaller in real
observations, challenging the detection of Stokes Q and U signals above
typical observational noise levels. A brief analysis of the effects of
noise is shown in the Appendix.

\section{Conclusions}\label{sec:conclusions}
We study the impact of finite spectral resolution on the retrieval of
magnetic field quantities from the Mg {\sc ii} h\&k lines through the
weak field approximation.

First, we look at the case of the WFA applied to synthetic Mg {\sc ii}
spectra emerging from a semi-empirical FAL-C model atmosphere with an
ad hoc uniform magnetic field with a LOS component of 141~G. Eq
\ref{eq:blos} is applied to the spectrally resolved case in two
wavelength ranges: one that contains only the inner lobes of Stokes V
and one that encompasses the outer lobes as well. We find that the
error in the retrieval of $B_{\rm LOS}$ jumps from 1.4\% in the case
of the narrower wavelength range, to -13\% when the outer lobes are
considered (left panels of Fig. \ref{fig:inner_outer_lobes}). The
results remain almost identical when the
spectra are synthesized without magneto-optical effects. It is
only when PRD is turned off that the WFA inferences in the two
spectral ranges agree and yield the smallest error (0.2\%), confirming
the results from \citet{hanlert}. It is well know than PRD effects
affect much more prominently the wings than the line core
\citep[][]{leenaarts1,leenaarts2}, and we speculate that the
  correction to the emissivity term due to PRD is purely responsible for the WFA
  breaking down.

Broad spectral point spread functions have the effect of spectral
mixing, bringing information from the line wings into the core and
vice-versa. We degrade the data spectrally by convolving them with
gaussian PSFs of different widths that emulate various spectral
resolutions. We find that, when we apply the WFA to the inner lobes of
Stokes V in the degraded spectra, a larger error ensues. By repeating
the experiment turning off PRD effects, we show that the latter are
purely responsible for the inference errors, as the spectral smearing
brings information from the line wings into the inner wavelength range
around the core of the line.
Fig. \ref{fig:WFA_with_R} shows the inaccuracy of the inferred
of $B_{\rm LOS}$ as a function of spectral resolution. The relative
error decreases with increasing spectral resolution, from $\sim 12$\% for
R = 30,000 to $\sim 3$\% for R = 90,000.

In order to test this effect in a more diverse set of atmospheres, we
synthesize the Mg {\sc ii} h\&k spectra emerging from a rMHD
simulation of an active region from the MURaM code. 
We select a portion of the simulation box that contains a quadrant of a sunspot
and the strongly magnetized granulation surrounding it.
When
Eq. \ref{eq:blos} is applied to the inner lobes of Stokes V, both Mg
{\sc ii} h and Mg {\sc ii} k deliver remarkably accurate results for
most pixels in the FoV, when comparing the retrieved value of the LOS
component of the magnetic field to the model value at $\tau_{\rm LINE}
= 1$. The WFA
inference starts to deviate significantly from the model value starting
at a magnetic field strength of $\sim 1500$~G (right panel of
Fig. \ref{fig:WFA_Blos_hvsk}), due to a departure from the weak field regime. Both spectral lines deliver quantitatively
comparable results. The effect of a reduced spectral resolution
results in the WFA underestimating the LOS magnetic field when
compared to the spectrally resolved case. We justify
that this bias is due to PRD effects being transported into the line core
due to spectral smearing, rather than a result of sensing lower
atmospheric layers due to line-of-sight mixing.

The extraction of the transverse component of the magnetic field
through Eq. \ref{eq:bt2} was tested
subsequently in the line core (defined here as the region spanned
by the inner lobes of Stokes V). We find that performing a linear
regression of the left and right handsides of the equation yields
better results when the intercept is forced to be zero than when it is
a free parameter of the fit (see first and second panels of the top row
of Fig. \ref{fig:WFA_Btrans_vs_R}). This is
because the WFA model does not reproduce the behavior of the data with
fidelity. A simple ratio of the wavelength averages of the left and right
sides of Eq. \ref{eq:bt2} (third panel of the top row) delivers
equally good results, while Eq. \ref{eq:marian} results in
systematically lower inferences of $B_{\rm T}$ (fourth panel of the
top row).
Degrading the spectral resolution results in a systematic
over-estimate of the retrieved $B_{\rm T}$, which grows with decreasing value of
$R$. The precision of the estimate also worsens (see the
increased spread in the scatter plots of
Fig. \ref{fig:WFA_Btrans_vs_R}), yielding method \#1 useless in the
case of R=30,000. For methods \#2 and \#3 , the systematic
offset of the WFA retrieval could be calibrated with a simple linear
fit.

One big caveat to the WFA is that the method does not shed information
on the atmospheric heights of the retrieved magnetic
quantities. The WFA inferences will correspond to weighted averages
around the height of formation of the spectral window that is used (typically the line core).
Across the domain of the MURaM simulation used in this
work, the geometric height of formation of the Mg {\sc ii} cores varies by
$\approx 1.4$~Mm, being very depressed at the center of the umbra and
a lot higher in the quieter areas. A similar variation is expected for
the heights probed by the WFA. This could pose a problem for
extrapolations of the vector magnetic field inferred from
observations, especially in the absence of further modeling that constrains the height probed
by the WFA.

 It is interesting to note that synthesizing the Mg {\sc ii}
   spectra in a simulation box featuring
11,250 1D realizations of the atmospheric properties took 145,000 core-hours on NCAR's Cheyenne
supercomputer. This highlights one of the difficulties of
interpreting the Mg {\sc ii} polarization spectra (and other
chromospheric resonance lines). The forward modeling alone is
computationally expensive, and traditional inversion methods based on
merit function minimization schemes tend to require many calls of the forward
model until the synthetic profiles converge to the observations,
limiting the use of this approach. That is why there are several
ongoing explorations of machine-learning based inversion algorithms,
or of model parameterizations that can significantly reduce the number
of forward-model iterations. Physical approximations such
as the WFA remain valuable to render the magnetic field retrieval tractable.

In conclusion, the WFA can be applied to the Mg {\sc ii} h\&k lines
emerging from strong field regions (sunspots and strongly magnetized surroundings),
but one has to be aware of the systematic impacts that finite spectral
resolution has on the inferences.

\appendix{}

 The magnetic field inferences through the WFA must be impacted by photon noise
in the observations, resulting not just in uncertainties, but also in
systematic biases \citep{marian2012, centenoWFA}. Even though it is beyond the
scope of this work to study the effects of noise in detail, we briefly
evaluate the relative merits of the different methods used in Section
\ref{sec:bt} for the inference of $B_{\rm T}$ when gaussian noise is
applied to the spectra.
To quantify the effects of noise we add wavelength-independent gaussian noise to Stokes $I$, $Q$, $U$
and $V$ and repeat the WFA inferences for the spectrally resolved
case. Three different signal-to-noise ratios (SNR) are considered,
$SNR = I_{\rm C} / \sigma = [\infty, 1000, 100]$, where the
reference signal, $I_{\rm C}$, is the
pseudo-continuum intensity 3.3~\AA\ blue-wards of Mg {\sc ii} k, and
$\sigma$ is the standard deviation of the noise.

\begin{figure}[!t]
  \includegraphics[angle=0,width=0.98\textwidth]{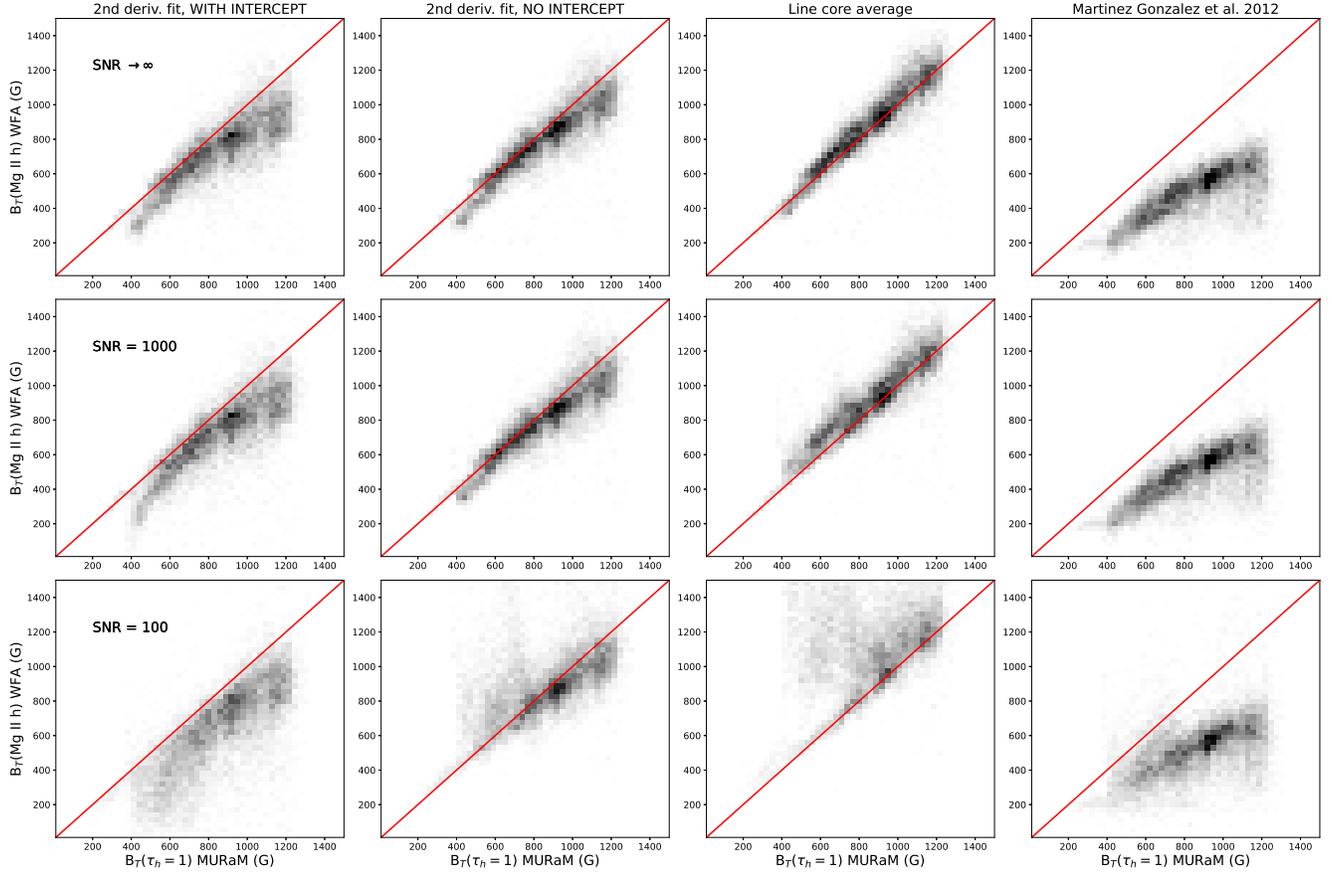}
  \caption{Scatter density plots of the values of B$_{\rm
      T}$ retrieved from Mg {\sc ii} h against their model
    counterparts for three different SNR values: SNR $\rightarrow
    \infty$ (top row), SNR = 1000 (middle row) and SNR = 100
    (bottom row). The four columns correspond to retrieval methods \#1, \#2, \#3
    and \#4 respectively, described in the manuscript text. The darker the grey-level, the higher the number of
    samples in the bin, which is 40~G in width. The color scale is kept constant across all panels. The red line represents the
    one-to-one correspondence. 
\label{fig:noise}}
\end{figure}

Figure \ref{fig:noise} shows the WFA inferences of $B_{\rm T}$ from Mg
{\sc ii} h for the
noiseless case in the top row (which is the same as the top row of
Fig.\ref{fig:WFA_Btrans_vs_R}), for $SNR = 1000$ in the middle
row, and $SNR = 100$ in the bottom row. While for a $SNR = 1000$ the
results are still robust, a $SNR=100$ not only increases the spread of the inferences around the ideal
solution, but also introduces systematic biases in almost all of the cases.
While method \#1 (linear fit of Eq. \ref{eq:bt2} with free-fitting
intercept) tends to experience a systematic bias towards lower
inferred values of $B_{\rm T}$, methods \#2 and \#3 are affected in
the opposite direction, with a tendency to overestimate $B_{\rm
  T}$. Method \#4 also experiences a bias towards larger inferred
values, but is less dramatically affected than the previous two methods.

The inference of $B_{\rm LOS}$ is not affected by the noise levels
tested above. Because the amplitude of Stokes $V$ across the FoV of
the simulation domain is large, even a SNR=100 renders circular
polarization signals significantly above the noise level, retaining
their diagnostic power with little to no negative impacts.

\clearpage

This material is based upon work supported by the National Center for
Atmospheric Research, which is a major facility sponsored by the
National Science Foundation under Cooperative Agreement
No. 1852977.
We would like to acknowledge high-performance computing support from
Cheyenne (doi:10.5065/D6RX99HX) provided by NCAR's Computational and
Information Systems Laboratory, sponsored by the National Science
Foundation.
T.d.P.A. acknowledges the funding received from the European Research Council (ERC) under the European Union’s Horizon 2020 research and innovation programme (ERC Advanced Grant agreement No 742265).

\bibliography{centeno.bib}{}
\bibliographystyle{aasjournal}

\end{document}